\documentclass[showpacs,aps,prl,reprint,superscriptaddress,preprintnumbers,nolinenumbers]{revtex4-1}

\usepackage{amsmath}
\usepackage{amssymb}
\usepackage{graphicx}
\usepackage{dcolumn}
\usepackage{color}
\usepackage{xcolor}
\usepackage{endnotes}
\usepackage{bm}
\usepackage{epsfig}
\usepackage{array}
\usepackage{appendix}
\usepackage{balance}

\newcommand{\etal}{{\it et al.}}
\newcommand{\ie}{{\it i.e.\ }}
\newcommand{\FTS}{{Fe$_{0.05}$TaS$_2$}}

\begin{document}



\title{Stabilization of antiferromagnetism in 1T-Fe$_{0.05}$TaS$_2$ }

\author{Q.~Niu$^\ddagger$}
\author{W.~Zhang$^\ddagger$}
\affiliation{Department of Physics, The Chinese University of Hong Kong, Shatin, Hong Kong}
\author{Y.~T.~Chan}
\affiliation{Department of Physics, The Chinese University of Hong Kong, Shatin, Hong Kong} 
\author{E.~C.~T.~O'Farrell}
\author{R.~Doganov}
\affiliation{Department of Physics, National University of Singapore, Singapore}
\author{K.~Y.~Yip}
\author{Kwing~To~Lai}
\affiliation{Department of Physics, The Chinese University of Hong Kong, Shatin, Hong Kong} 
\author{W.~C.~Yu}
\affiliation{Department of Physics, City University of Hong Kong, Kowloon, Hong Kong}
\author{B.~\"{O}zyilmaz}
\affiliation{Department of Physics, National University of Singapore, Singapore}
\author{G.~R.~Stewart}
\author{J.~S.~Kim}
\affiliation{Department of Physics, University of Florida, Gainesville, Florida 32611-8440, USA}
\author{Swee~K.~Goh}
\email[]{skgoh@cuhk.edu.hk}
\affiliation{Department of Physics, The Chinese University of Hong Kong, Shatin, Hong Kong}

\date{\today}

\begin{abstract}
1T-TaS$_2$ is a prototypical charge-density-wave (CDW) system with a Mott insulating ground state. Usually, a Mott insulator is accompanied by an antiferromagnetic state. However, the antiferromagnetic order had never been observed in 1T-TaS$_2$. Here, we report the stabilization of the antiferromagnetic order by the intercalation of a small amount of Fe into the van der Waals gap of 1T-TaS$_2$, \ie\ forming 1T-Fe$_{0.05}$TaS$_2$. Upon cooling from 300~K, the electrical resistivity increases with a decreasing temperature before reaching a maximum value at around 15~K, which is close to the Neel temperature determined from our magnetic susceptibility measurement. The antiferromagnetic state can be fully suppressed when the sample thickness is reduced, indicating that the antiferromagnetic order in \FTS\ has a non-negligible three-dimensional character. For the bulk \FTS, a comparison of our high pressure electrical transport data with that of 1T-TaS$_2$ indicates that, at ambient pressure, Fe$_{0.05}$TaS$_2$ is in the nearly commensurate charge-density-wave (NCCDW) phase near the border of the Mott insulating state. 
The temperature-pressure phase diagram thus reveals an interesting decoupling of the antiferromagnetism from the Mott insulating state.\\

\end{abstract}


\maketitle

\section{Introduction}
Mott insulators are an important class of materials in strongly correlated electron research. The sustained interest in Mott insulators can be largely attributed to the fact that several distinct families of superconductors are found in their vicinity ({\it e.g.} Refs.~\cite{Lee2006, Song2016, Sipos2008, Nam2007, Cao2018a, Cao2018b}). In these systems, the superconducting phase arises when the Mott insulating state is tuned away either by controlling the band filling or by varying the bandwidth. For instance, a Mott insulating phase in the twisted bilayer graphene can be driven into a superconducting state by electrostatic doping the bilayer graphene at a magic twist angle \cite{Cao2018a, Cao2018b}.

1T-TaS$_2$ is an interesting layered transition metal dichalcogenide featuring a series of charge-density-wave (CDW) transitions \cite{Friend1987, Wilson1975,Fazekas1980,Fazekas1979}. At very high temperature ($T>550$~K), 1T-TaS$_2$ is metallic. Upon cooling, 1T-TaS$_2$ enters an incommensurate CDW (ICCDW) phase, a nearly commensurate CDW (NCCDW) phase, and a commensurate CDW (CCDW) phase at 550~K, 350~K and 180~K, respectively. In the CCDW phase, the displacement of the Ta atoms in the layer forms clusters of David-stars, in which one Ta atom is located at the centre of each David-star formed by 12 Ta atoms. Thus, there is one unpaired 5$d$ electron in each cluster which, according to band theory, should give rise to a metallic state. Therefore, the observed insulating behaviour in the CCDW phase must be due to a Mott insulating state driven by strong electron-electron correlations \cite{Fazekas1979}. 

Various efforts have been devoted to tune the phases in 1T-TaS$_2$ \cite{Sipos2008, Li2012, Ang2012, Yu2015, Ang2013, Wang2018}. Under pressure \cite{Sipos2008,Wang2018}, the CCDW/Mott state first disappears at around 10~kbar. Superconductivity appears at $\sim$20~kbar. Concomitantly, the NCCDW phase is rapidly suppressed. Interestingly, the superconducting transition temperature ($T_c$) does not vary significantly after reaching the maximum value at around 40~kbar. This insensitivity to pressure despite a vastly different normal state suggests the irrelevance of superconductivity to CDW fluctuations. Nevertheless, 1T-TaS$_2$ is another example in which the interplay between superconductivity and the Mott state can be recognized.

A long-standing puzzle in the studies of 1T-TaS$_2$ concerns the nature of the magnetism associated with the CCDW/Mott state. Conventionally, a Mott insulator is accompanied by antiferromagnetism (AFM), which has never been conclusively observed in 1T-TaS$_2$. An indirect evidence of the antiferromagnetic order was provided by an angle-resolved photoemission measurement, which detected an unexpected periodicity in the quasiparticle dispersion \cite{Perfetti2005}.  Furthermore, even in the absence of a magnetic order, the local moments due to Ta atoms at the centre of David-stars should result in Curie-Weiss temperature dependence of the magnetic susceptibility. Although such behaviour has indeed been observed recently \cite{Ribak2017, Kratochvilova2017}, the Curie constant is rather small, indicating a very small concentration of local moments. To explain the enigmatic magnetic properties, Law and Lee proposed that 1T-TaS$_2$ is a quantum spin liquid \cite{Law2017}. Subsequent thermal conductivity measurements show contradictory results on the existence of the residual linear term at zero field, which is sensitive to itinerant magnetic excitations \cite{Yu2017, Murayama2019}

The layered nature of 1T-TaS$_2$ enables the intercalation of atoms or molecules into the van der Waals gap to tune the material properties \cite{Friend1987}. In this work, we describe our success to intercalate Fe atoms with a nominal concentration of 5\% per formula unit to the 1T phase of TaS$_2$, \ie\ forming 1T-\FTS. We show that the ground state of 1T-\FTS\ is antiferromagnetic. Further analysis of our high pressure transport data shows that we can construct a universal temperature-pressure phase diagram for 1T-TaS$_2$ and 1T-\FTS\ where we find that the antiferromagnetism is detached from the Mott state.


\section{Methods}
Single crystals of \FTS\ were grown using the iodine chemical vapour transport method, 
with the hot end of the quartz tube at 1000$^\circ$C and the cold end at 900$^\circ$C.  The purities of the elemental powders, in stoichiometric amounts, placed at the hot end were Ta:  99.98\%, S:  99.999\% (metals basis), Fe:  99.998\%. The intercalation nature of our crystals was confirmed by energy-dispersive X-ray spectroscopy, as discussed in the appendix. The magnetic susceptibility was measured using a SQUID magnetometer by Quantum Design. High pressure electrical resistivity was measured by a standard four-terminal configuration using a piston-cylinder clamp cell in a Physical Property Measurement System by Quantum Design or a dilution refrigerator by Bluefors. Glycerin was used as the pressure transmitting medium and the pressure value was obtained by measuring the zero-field superconducting transition of a piece of Pb next to the sample. Thickness dependent resistivity measurements were made by mechanical exfoliation of crystals onto highly doped Si/SiO$_2$ substrates. The thickness of individual flakes was determined by atomic force microscope and selected flakes were then contacted using standard lithographic techniques with Cr/Au contacts. For the density functional theory calculations, the optimized lattice structural parameters were obtained using the Quantum Espresso package \cite{Giannozzi2009} with ultrasoft pseudopotential and Perdew-Burke-Ernzerhof (PBE) \cite{Perdew1996} exchange-correlation density functional. The plane-waves kinetic energy cutoff of 50 Ry and a $k$-mesh of $25\times25\times12$ were used. The electronic band structure calculations were performed using the all-electron full-potential linearized augmented plane-wave code WIEN2k \cite{Schwarz2003}. The parameter $R_{\rm MT}^{\min}K_{\max}$ was set to 7 and a $k$-point mesh of 8000 in the first Brillouin zone was used.
 
\section{Results and discussions}
\begin{figure}[!t]\centering
      \resizebox{8.5cm}{!}{
              \includegraphics{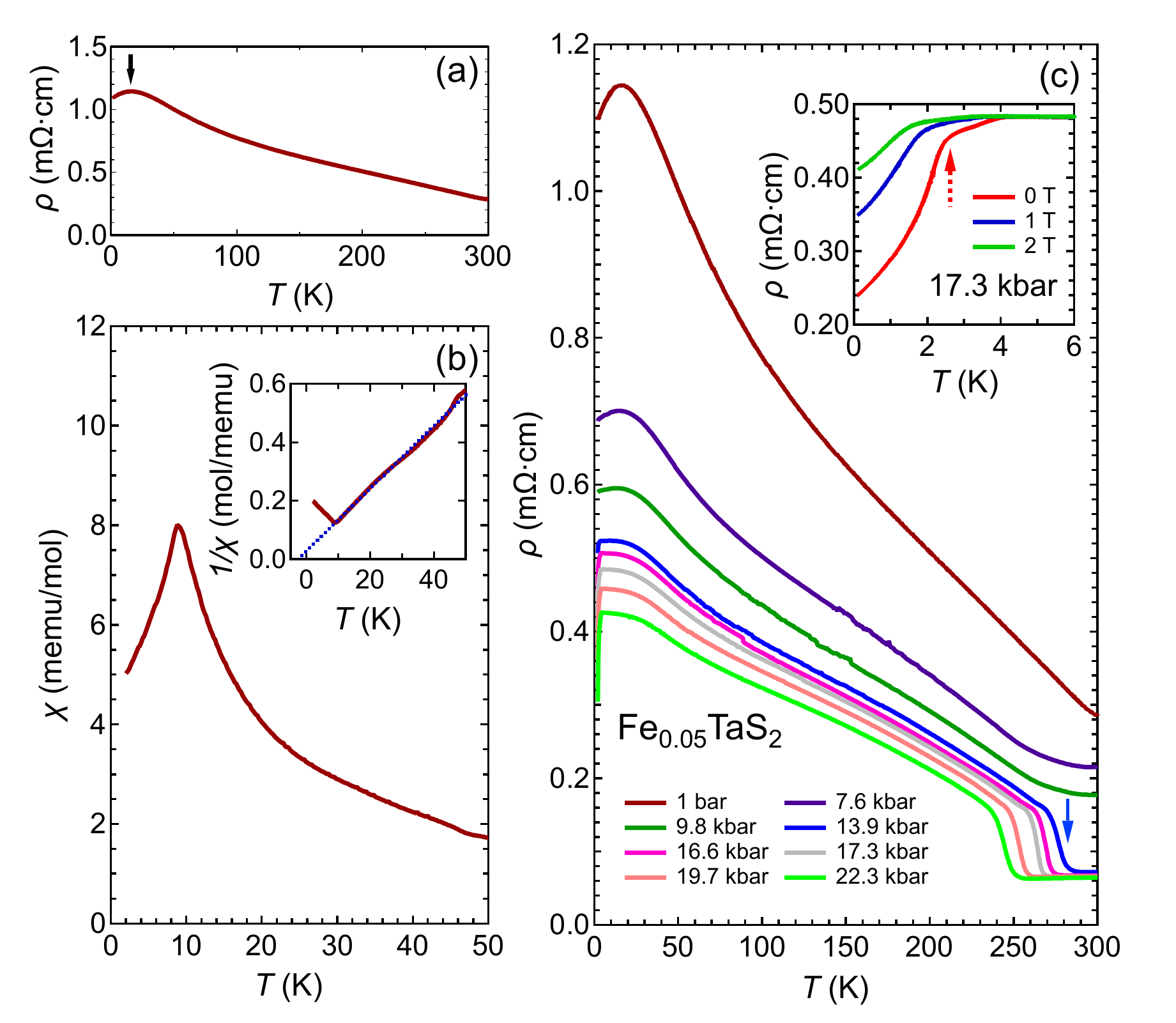}}                				
              \caption{\label{fig1} (a) Temperature dependence of the electrical resistivity in \FTS\ at ambient pressure and zero field. The vertical arrow denotes the peak temperature of 15.7~K. (b) Temperature dependence of the magnetic susceptibility in \FTS. The data were collected with 50 Oe applied along the $c$ axis. The inset shows the same dataset plotted as the inverse of the magnetic susceptibility against the temperature. The dotted line is the linear fit of the data from 9.2~K to 50~K. (c) Temperature dependence of the electrical resistivity in \FTS\ under pressure. The data were collected on cooling. The vertical arrow indicates $T_{\rm NCCDW}$ at 13.9~kbar, defined as the temperature at which $d\rho/dT$ is a local minimum. The inset shows the low temperature region of $\rho(T)$ at 17.3 kbar, collected at 0~T, 1~T and 2~T. The field is along the $c$ axis. $T_{c, \rm onset}$ is defined as the temperature below which $\rho(T)$ drops substantially (see the dashed vertical arrow).}
\end{figure}

Figure~\ref{fig1}(a) shows the temperature ($T$) dependence of the electrical resistivity ($\rho$) in \FTS\ at ambient pressure. $\rho(T)$ increases monotonically on cooling from 300~K to $\sim$15.7~K, before decreasing below $\sim$15.7~K. This peak-like structure is not seen in the pristine TaS$_2$ at ambient pressure, in which $\rho(T)$ experiences a much stronger enhancement over several orders of magnitude upon cooling from 300~K to 2~K (see {\it e.g.}, \cite{Sipos2008,Wang2018,Ribak2017}). Figure~\ref{fig1}(b) displays the $T$ dependence of the magnetic susceptibility ($\chi$) in \FTS\ measured in the zero-field cooled mode. Interestingly, $\chi(T)$ exhibits a peak at $\sim$8.9~K, characteristic of an antiferromagnetic order. Such a peak-like structure in $\chi(T)$ is also not seen in the pristine TaS$_2$, in which $\chi(T)$ follows the Curie-Weiss law with a very small Weiss temperature down to the lowest temperature \cite{Ribak2017}. From both $\rho(T)$ and $\chi(T)$, we conclude that the ground state of \FTS\ is antiferromagnetic, which is distinct from that of the pristine TaS$_2$, and we attribute the peak in $\rho(T)$ to the onset of the antiferromagnetic order.
\begin{figure}[!t]\centering
      \resizebox{8.5cm}{!}{
              \includegraphics{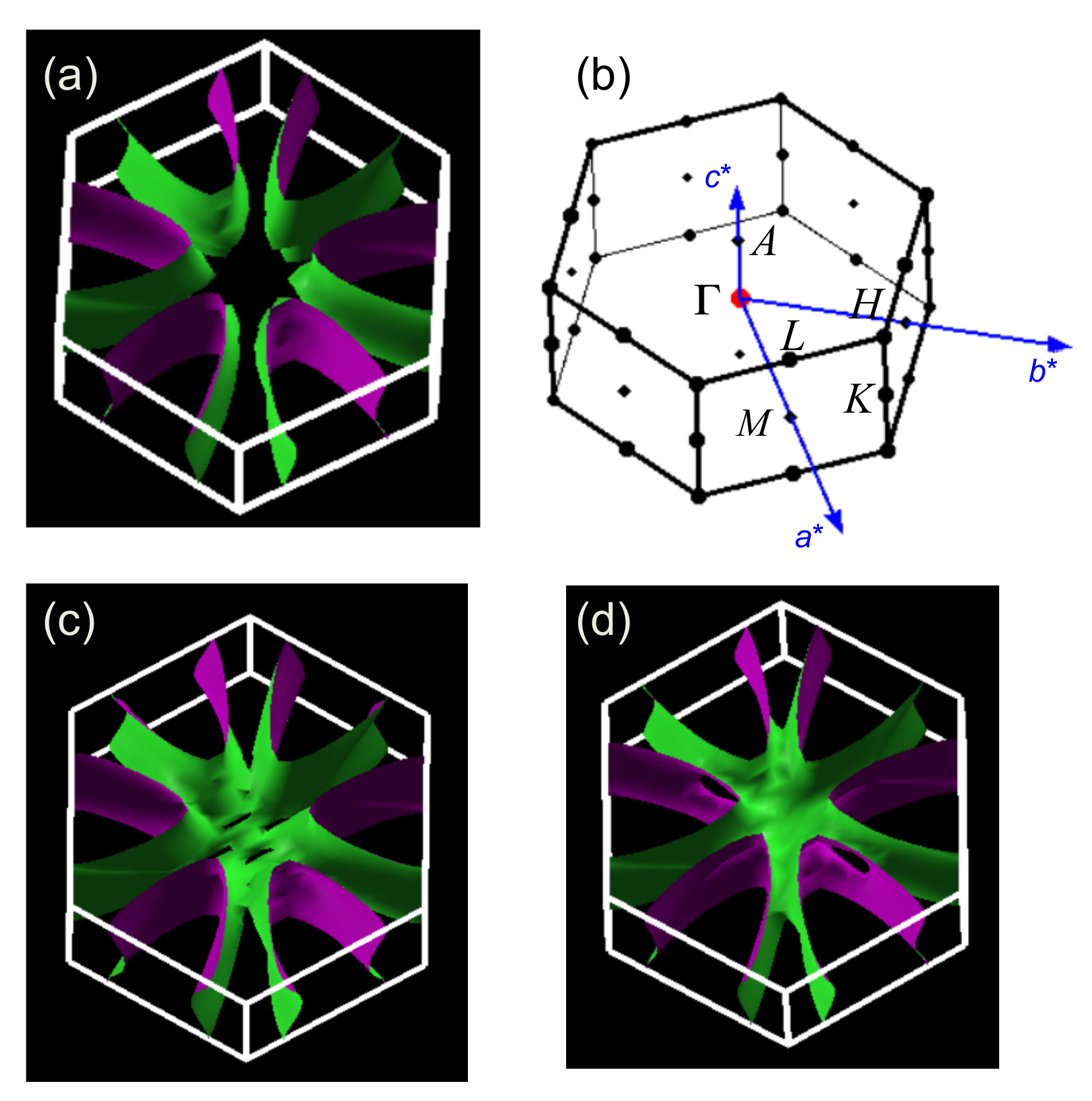}}            			
              \caption{\label{fig2} Calculated Band 20 Fermi surface of the undistorted 1T-TaS$_2$ (a) at ambient pressure, (c) at 20 kbar, and (d) when the Fermi energy is shifted upward by 0.01~Ry relative to the band structure at ambient pressure, {\it i.e.}, $E_F=E_F(p=0)+0.01$~Ry. The first Brillouin zone and the high symmetry points of 1T-TaS$_2$ are shown in (b), and $a^*$, $b^*$ and $c^*$ are the basis vectors of the reciprocal lattice. $c^*$ is parallel to the $k_z$ axis.
              }            
\end{figure}

As mentioned in the Introduction, a central issue in the study of the pristine TaS$_2$ concerns the absence of an antiferromagnetic order. In addition, an analysis of magnetic susceptibility data of the pristine TaS$_2$ revealed an effective moment that has been attributed to a very small concentration of Ta spins. Assuming a $g$-factor of 2, Ribak \etal\ estimates that only 1 out of 2450 Ta moments contributes to the magnetic susceptibility \cite{Ribak2017}. Because our magnetic susceptibility data above 9.0~K also follow a Curie-Weiss law $\chi=C/(T-\theta_{cw})$, we can estimate the effective moment in the paramagnetic state of \FTS\ through the Curie constant $C$. In the inset of Fig.~\ref{fig1}(b), we plot $\chi^{-1}$ against $T$, from which we determine the Curie constant $C=9.35\times10^{-2}$~emu\ K/mol and the Weiss temperature $\theta_{cw}=-2.55~$K. We obtain an effective moment $\mu_{\rm eff}=0.865\mu_{\rm B}/$f.u. from the Curie constant. This effective moment is significantly larger than the value of 0.035$\mu_{\rm B}$/f.u. estimated from the Curie constant of 1T-TaS$_2$ reported by Ribak \etal\ \cite{Ribak2017}.

To gain further insight, it is instructive to compare the relative position of \FTS\ on the $T$-$p$ phase diagram of the pristine 1T-TaS$_2$. Figure~\ref{fig1}(c) shows $\rho(T)$ of \FTS\ under hydrostatic pressure. The overall resistivity decreases with an increasing pressure. At 7.6~kbar and 9.8~kbar, the $\rho(T)$ traces are qualitatively similar to the ambient pressure data, apart from a flattening of the peak at $\sim$15~K and a change of curvature near 300~K. At 13.9~kbar, the resistivity exhibits a step-like increase upon cooling across 280~K. Furthermore, a downturn begins to develop near the base temperature of our measurement. Both features are sensitive to applied pressure. With a further increase of pressure, the step-like increase occurs at a lower temperature, while the downturn becomes more pronounced. 

Comparing with the ambient pressure resistivity data of 1T-TaS$_2$, the step-like increase of $\rho(T)$ in \FTS\ can be attributed to a phase transition temperature from an ICCDW state to a NCCDW state ($T_{\rm NCCDW}$), while the low-temperature downturn that begins to develop at high pressures can be regarded as signaling a superconducting transition. The inset of Figure~\ref{fig1}(c) shows the low temperature region of $\rho(T)$ at 17.3 kbar. The downturn can be suppressed when a magnetic field is applied. Thus, this reinforces the view that the downturn is related to superconductivity despite the absence of a zero resistance.

From Fig.~\ref{fig1}(c), we can determine the pressure dependence of $T_{\rm NCCDW}$, defined as the temperature at which $d\rho/dT$ is a local minimum.  Interestingly, $T_{\rm NCCDW}$ in \FTS\ is suppressed at the rate of $-3.7$~K/kbar, which is the same as that in the pristine 1T-TaS$_2$ \cite{Sipos2008}. Analysis of the pressure dependence of $T_{\rm NCCDW}$ suggests that the intercalation of 5\% Fe is equivalent to applying $\sim$13~kbar to the pristine 1T-TaS$_2$. With this information, we construct the $T$-$p$ phase diagram (Fig.~\ref{fig3}), where the bottom axis is the pressure scale for 1T-TaS$_2$ while the top axis is the pressure scale with an offset of 13~kbar for \FTS.

The observation that the effect of Fe intercalation is similar to applied pressure is also supported by our density functional theory calculations. Figure~\ref{fig2}(a) shows the calculated Band 20 Fermi surface of the pristine 1T-TaS$_2$ at ambient pressure, where six elliptical electron cylinders are centered around the edges of the first Brillouin zone. The high symmetry points in the first Brillouin zone are shown in Fig.~\ref{fig2}(b). When pressure is applied, the elliptical cylinders expand and eventually touch each other, resulting in a Fermi surface with a normal vector along the $k_z$ direction at the zone center ($\Gamma$) (Fig.~\ref{fig2}(c)). The effect of Fe intercalation is usually understood in the `rigid-band approximation', in which the intercalant is localized and donates charge to the host whose band structure is unchanged \cite{Friend1987}. Therefore, we can investigate the effect of the Fe intercalation, which introduces electrons \cite{Friend1987}, by raising the Fermi energy with respect to the band structure of 1T-TaS$_2$ at ambient pressure. A very similar Fermi surface can be obtained (Fig.~\ref{fig2}(d)). Thus, the evolution of the electronic structure when Fe is intercalated is similar to applying pressure on the pristine TaS$_2$.

The $T$-$p$ phase diagram constructed reveals an antiferromagnetic region which is separated from the CCDW/Mott phase. We point out that the peak in $\rho(T)$ shown in Fig.~\ref{fig1}(a) was also observed in 1T-TaS$_2$ at around 10~kbar, as displayed in Fig.~2(a) of Ref.~\cite{Sipos2008}, although the feature was not discussed. Therefore, it is conceivable that a similar antiferromagnetic region has also been stabilized in 1T-TaS$_2$ under pressure. On the phase diagram, the onset temperature of the superconductivity $T_{c, \rm onset}$ in \FTS\ is also included. $T_{c, \rm onset}$ increases slightly with an increasing pressure, morphing smoothly into the maximum $T_c$ region of 1T-TaS$_2$. 

\begin{figure}[!t]\centering
       \resizebox{8.5cm}{!}{
              \includegraphics{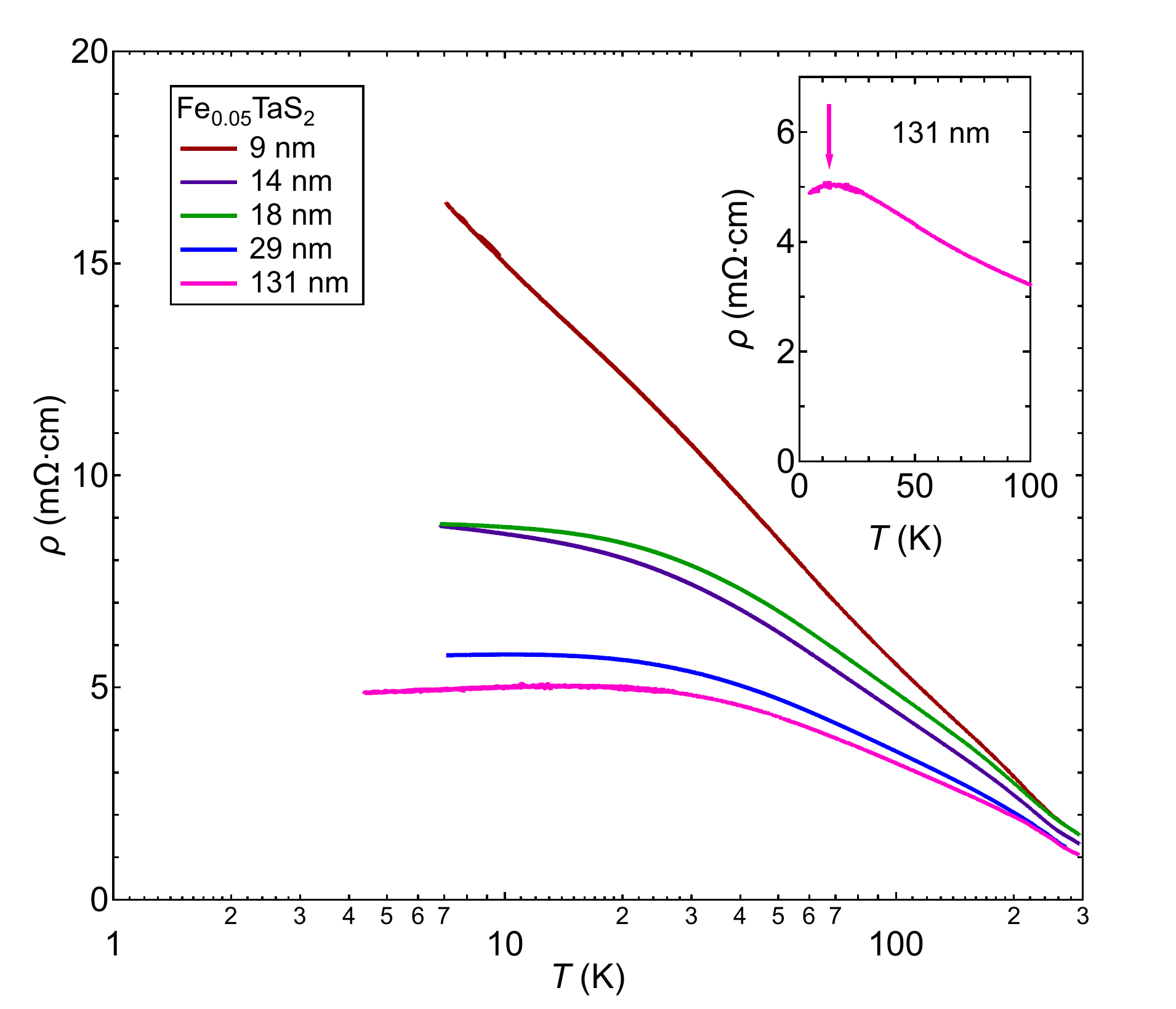}}                				
              \caption{\label{fig4} Temperature dependence of the electrical resistivity in \FTS\ flakes with different thicknesses ranging from 9~nm to 131~nm. Note the logarithmic scale for the temperature axis. Inset: the low temperature part of the resistivity in the \FTS\ flake with the thickness of 131~nm, where a peak similar to the data on the bulk sample is observed, as indicated by the arrow.}
\end{figure}

Recently, magnetism has been stabilized in systems such as CrI$_3$ \cite{Huang2017} and Fe$_3$GeTe$_2$ \cite{Deng2018} even when they approach the monolayer limit. To further examine the thickness dependence of the antiferromagnetic state in \FTS, we fabricate devices using flakes of \FTS\ with different thicknesses. Figure~\ref{fig4} displays $\rho(T)$ for samples with different thicknesses. For the flake which is 131~nm thick, a similar peak in $\rho(T)$ is observed at 12.9~K (see also the inset). However, when the thickness is reduced, the peak gradually disappears. Thus, the sample with a thickness of 131~nm can be regarded as in the bulk limit. At 9~nm, a striking $\log(T)$ dependence of the resistivity is instead observed from 300~K down to the lowest temperature measured. From this dataset, we conclude that approaching 2D limit is detrimental to antiferromagnetism, and the antiferromagnetic order has a substantial 3D character. This behaviour of antiferromagnetism is reminiscent of the case of dimensional tuning in CeIn$_3$/LaIn$_3$ superlattices \cite{Shishido2010}.

We now attempt to understand the source of the effective moment in \FTS\ obtained from the Curie-Weiss analysis mentioned above. The first candidate is Fe. In TaS$_2$, Fe is usually localized and enters as Fe$^{2+}$ \cite{Ko2011, Eibschutz1980, Narita1994}. Thus, the spin-only effective moment due to Fe$^{2+}$ is $4.90\mu_{\rm B}$/Fe or $0.25\mu_{\rm B}$/f.u., which is smaller than the experimentally determined value of $0.865\mu_{\rm B}$/f.u. The second candidate is Ta$^{4+}$, each contributes one 5$d$ electron. For Ta$^{4+}$, the expected effective moment is 1.73$\mu_{\rm B}$/f.u., which is larger than our experimental result.   
\begin{figure}[!t]\centering
       \resizebox{8.5cm}{!}{
              \includegraphics{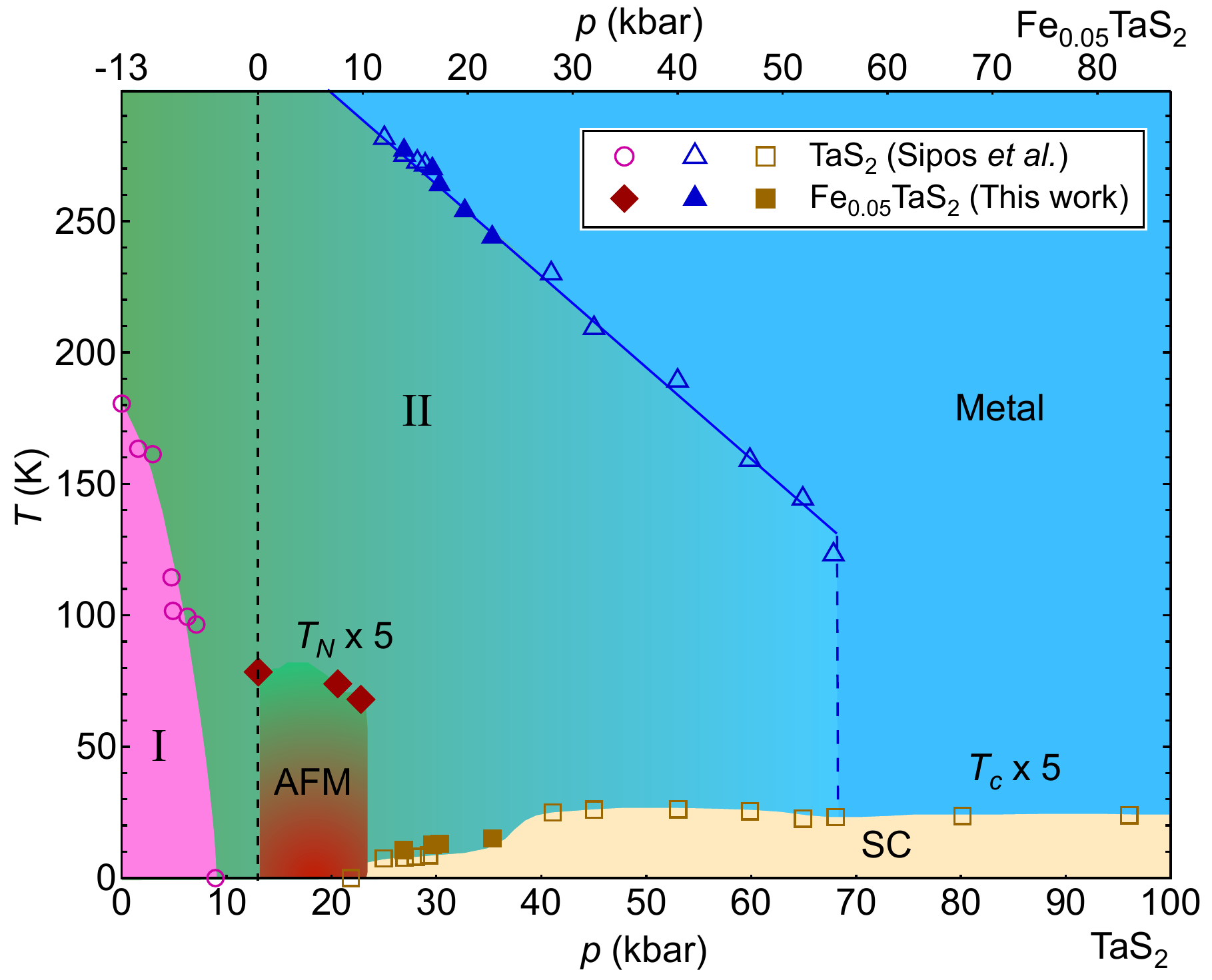}}                				
              \caption{\label{fig3} Temperature-pressure phase diagram for 1T-TaS$_2$ (bottom axis) and 1T-\FTS\ (top axis). The top axis is offset by 13~kbar relative to the bottom axis. The open symbols are from Ref.~\cite{Sipos2008}, and the closed symbols are data from this work. ``I" denotes the CCDW and the Mott state, ``II" denotes the NCCDW state.}
\end{figure}

Nevertheless, it is possible to understand the second scenario by considering the formation of David-stars in TaS$_2$. In the pristine TaS$_2$, when Ta atoms form David-stars, only the magnetic moment of the Ta atom at the center of each 12-atom star contributes to the magnetic susceptibility. Hence, if all Ta atoms participate in the formation of David-stars, as in the case of the CCDW phase of the pristine TaS$_2$, the effective moment is expected to be reduced by a factor of 13, \ie $0.13\mu_{\rm B}$/f.u. In \FTS\ at ambient pressure, the system is in the NCCDW instead of the CCDW phase. Hence, some Ta atoms do not participate in the formation of David-stars, resulting in a less substantial reduction of the effective moment. Although microscopic probes are needed to settle the origin of the magnetic moment, our analysis allows us to tentatively attribute the measured effective moment in \FTS\ to Ta atoms.

The properties of 1T-TaS$_2$ can also be tuned by the substitution of Ta by Fe, \ie\ forming the Fe$_y$Ta$_{1-y}$S$_2$ compounds \cite{Li2012, Ang2012}. The CCDW/Mott state can be fully suppressed with a small amount of Fe ($y=0.01$), and the superconductivity can be induced at $y=0.02$ and $y=0.03$, both in the NCCDW phase. At $y=0.04$, superconductivity disappears. At $y=0.05$, the electrical resistivity increases anomalously at low temperatures, which has been attributed to the Anderson localization \cite{Li2012}. The phases resulted from the Fe doping are clearly different from the present case of Fe intercalation, in which such an increase in the resistivity is not observed in \FTS. Thus, 1T-\FTS\ can provide a model system to further explore the magnetism and its interplay with neighbouring phases.

The $T$-$p$ phase diagram constructed in Fig.~\ref{fig3} displays an extraordinary decoupling of the antiferromagnetism from the Mott insulating state. Conventional wisdom dictates the coexistence of these phases, because the electronic system can gain energy by arranging the adjacent spins in the anti-parallel manner. Thus, it is urgently needed to understand the mechanism that drives the separation of these phases. Finally, the role of the antiferromagnetic order and its fluctuations might be important for the complete understanding of the superconductivity in this system.

\section{Conclusions}
In summary, we have synthesized and studied single crystals of 5\% Fe intercalated 1T-TaS$_2$, \FTS. An antiferromagnetic phase is detected in \FTS, which can be followed to $\sim$9.8~kbar. The thickness dependent resistivity measurements show that the antiferromagnetic order disappears with a reduction of the sample thickness. The analysis of our high pressure resistivity data demonstrates that \FTS\ is located at $\sim$13~kbar with respect to the temperature-pressure phase diagram of the pristine 1T-TaS$_2$. With the inclusion of the \FTS\ data, the temperature-pressure phase diagram reveals an interesting decoupling of the antiferromagnetic region from the CCDW/Mott state, which warrants further theoretical and experimental studies. 

\section{Acknowledgments}
\begin{acknowledgments}
We acknowledge Antonio Castro Neto for discussions. The work was supported by Research Grants Council of Hong Kong (GRF/14300418, GRF/14300117), CUHK Direct Grant (4053345, 4053299), CityU Start-up Grant (No. 9610438), the National Research Foundation of Singapore under the Competitive Research Programme (Grant No. NRF-CRP9-2011-3), and the National Research Foundation of Singapore under the MediumSized Centre Grant. Work by J.S.K. and G.R.S. was supported by the US Department of Energy Basic Energy Sciences under Contract Nos. DE-SC-0020385 and DE-FG02-86ER45268.

$^\ddagger$Q.N. and W.Z. contributed equally to this work. 
\end{acknowledgments}



\section{Appendix: Energy-dispersive X-ray spectroscopy measurements}

\begin{figure}[!h]
	\includegraphics[width=.48\textwidth]{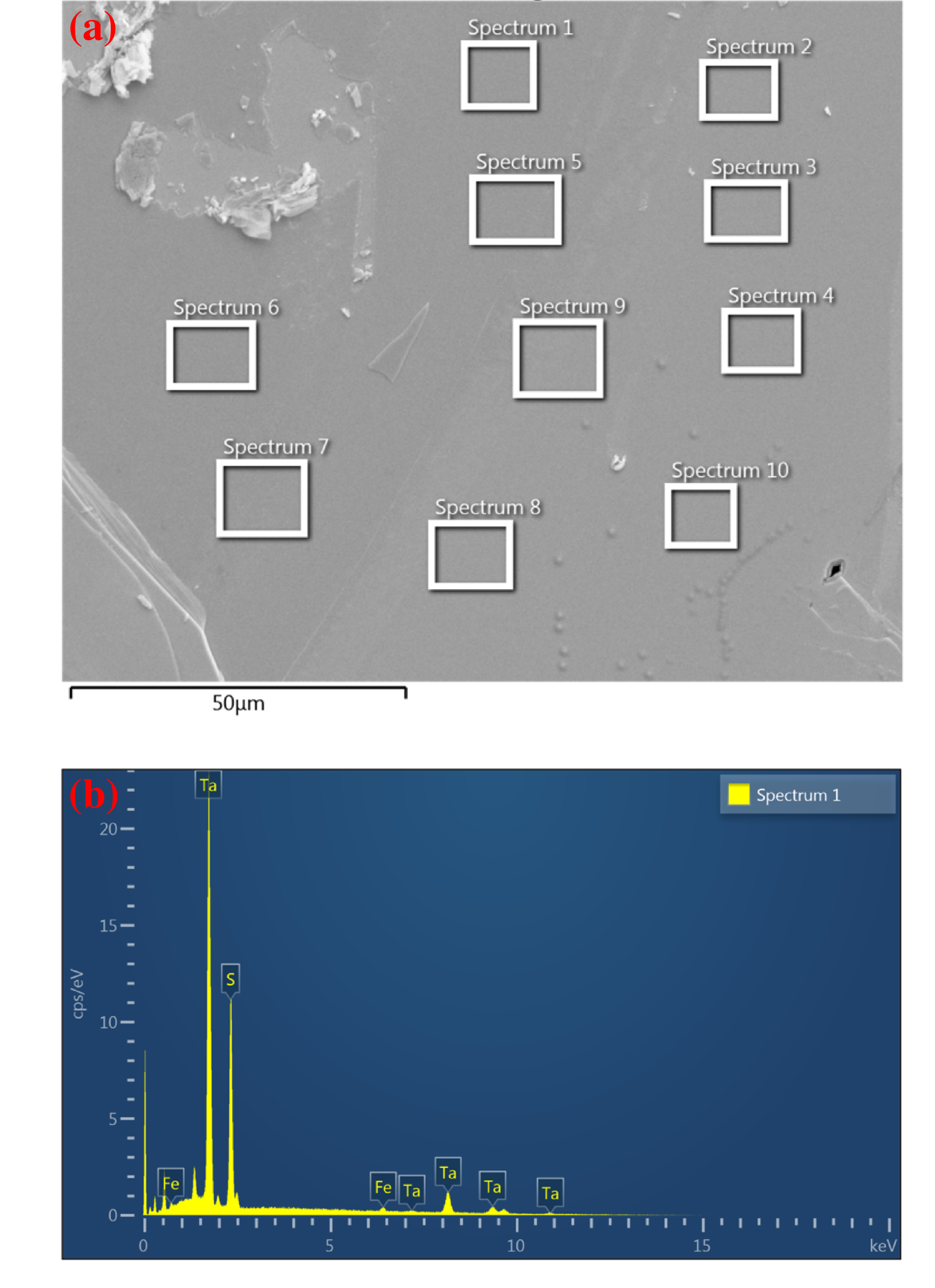}
	\caption{\label{fig5} (a) Scanning electron microscope image of Crystal~1. The white rectangles are the spots we chose to perform the measurements. (b) Energy-dispersive X-ray spectroscopy result of ``Spectrum 1" for \FTS~(Crystal~1). }
\end{figure}

In this section, we present energy-dispersive X-ray (EDX) spectroscopy results for \FTS. We measured three pieces of \FTS\ crystals. For each crystal, we performed the measurement at 10 random spots. Figure~\ref{fig5}(a) shows the spots for Crystal 1, and Fig.~\ref{fig5}(b) shows the spectrum corresponding to the spot labelled ``Spectrum 1". The relative atomic abundances of Fe, Ta and S are tabulated in Table~\ref{table1}, and each entry is the result of averaging the data over 10 spots. We set the atomic abundance of S to 2.000, and normalized the abundances of Fe and Ta to this value. Here, it is clear for the three different pieces that the ratio Ta:S is close to 1:2. If Fe atoms were substituting the Ta site, the relative abundance of Ta would be less than 1. This shows that Fe is an intercalant. 

The distribution of the Fe content across different crystals is probably responsible for the variation of the peak temperatures determined from $\rho(T)$ and $\chi(T)$.

\begin{table}[!h]
\caption{Summary of relative atomic abundances of Fe, Ta and S for three different \FTS\ samples.}
\label{table1}
\begin{center}
\begin{tabular}{|c | c | c | c | c | c | c | c | c |}
\hline
Sample                 & Fe                                        &Ta                                        &  S (as 2.000)       \\ 
\hline
~Crystal 1~        & ~0.076~$\pm$~0.013~       & ~1.046~$\pm$~0.014~      & ~2.000~$\pm$~0.019~      \\ 
\hline
~Crystal 2~        & ~0.087~$\pm$~0.010~       &  ~1.086~$\pm$~0.018~      &  ~2.000~$\pm$~0.014~   \\
\hline
~Crystal 3~        & ~0.094~$\pm$~0.008~       & ~1.107~$\pm$~0.010~      &  ~2.000~$\pm$~0.013~      \\ 
\hline

\end{tabular}
\end{center}
\end{table}


\balance
\providecommand{\noopsort}[1]{}\providecommand{\singleletter}[1]{#1}%

\end{document}